\documentclass[aps,showpacs,superscriptaddress,showpacs,twocolumn]{revtex4}
\usepackage[intlimits]{amsmath}
\usepackage{amssymb}
\usepackage{graphicx}
\usepackage{booktabs}
\usepackage{mathrsfs,slashed}
\usepackage{hyperref}
\usepackage{epsfig}
\usepackage{color}

\newcommand{\bea}{\begin{eqnarray}}
\newcommand{\eea}{\end{eqnarray}}
\newcommand{\beq}{\begin{equation}}
\newcommand{\eeq}{\end{equation}}

\newcommand{\ii}{{\rm i}}
\newcommand{\ee}{\,{\rm e}}

\DeclareMathOperator{\Tr}{Tr}
\DeclareMathOperator{\tr}{tr}

\begin{document}
\title{Polyakov-loop suppression of colored states in a
quark-meson-diquark plasma
}
\author{D.~Blaschke}
\email{blaschke@ift.uni.wroc.pl}
\affiliation{Instytut Fizyki Teoretycznej, Uniwersytet
Wroc{\l}awski, 50-204 Wroc{\l}aw, Poland} \affiliation{Bogoliubov
Laboratory for Theoretical Physics, JINR Dubna, 141980 Dubna,
Russia}

\author{A.~Dubinin}
\email{aleksandr.dubinin@ift.uni.wroc.pl}
\affiliation{Instytut Fizyki Teoretycznej, Uniwersytet
Wroc{\l}awski, 50-204 Wroc{\l}aw, Poland}

\author{M.~Buballa}
\email{michael.buballa@physik.tu-darmstadt.de}
\affiliation{Institute for Nuclear Physics, Technical University of Darmstadt,
	Schlossgartenstr. 2, 64289 Darmstadt, Germany}

\begin{abstract}
A quark-meson-diquark plasma is considered within the PNJL model for
dynamical chiral symmetry breaking and restoration in quark matter.
Based on a generalized Beth-Uhlenbeck approach to mesons and diquarks we
present the thermodynamics of this system including the Mott dissociation of
mesons and diquarks at finite temperature.
A striking result is the suppression of the diquark abundance below the chiral
restoration temperature by the coupling to the Polyakov loop, because of their
color degree of freedom.
This is understood in close analogy to the suppression of quark
distributions by the same mechanism.
Mesons as color singlets are unaffected by the Polyakov-loop suppression.
At temperatures above the chiral restoration mesons and diquarks are both suppressed
due to the Mott effect, whereby the positive resonance contribution to the
pressure is largely compensated by the negative scattering contribution in
accordance with the Levinson theorem.
\end{abstract}
\maketitle
\section{Introduction}			
For a quantum field theoretic description of hadronic correlations in quark matter that shares the  property of approximate chiral symmetry with the QCD Lagrangian, the Nambu--Jona-Lasinio (NJL) model has been widely used.
This model is particularly suitable to address the appearance of pions in quark matter as a consequence of dynamical chiral symmetry breaking in accordance with the Goldstone theorem. The absence of confinement is one of the shortcomings of the NJL model (see, e.g., 
\cite{Volkov:1984kq,Vogl:1991qt,Klevansky:1992qe,Hatsuda:1994pi} for early reviews and \cite{Buballa:2003qv,Fukushima:2013rx} for more recent ones with emphasis on the high density aspects).

For a solution of this problem it has been proposed to take into account the suppression of colored quark states by their coupling to the Polyakov loop.
For the Polyakov-loop extended NJL (PNJL) model \cite{Meisinger:1995ih,Fukushima:2003fw,Megias:2004hj,Ratti:2005jh} in the mean field approximation a quark distribution function arises which in the limit of the confining phase (where for the traced Polyakov loop holds $\Phi \to 0$) is strongly suppressed
relative to the ordinary Fermi function for quarks obtained in the deconfined phase where $\Phi \to 1$.
Indeed this has been demonstrated in Refs. \cite{Fukushima:2003fw,Ratti:2005jh}. The model has been extended to also include mesonic correlations, see  \cite{Hansen:2006ee,Schaefer:2007pw,Blaschke:2007np,Rossner:2007ik,Radzhabov:2010dd,Horvatic:2010md}. Note that the coupling to the Polyakov-loop also suppresses thermodynamic instabilities in nonlocal NJL models  \cite{Benic:2012ec,Benic:2013eqa}.

In the present work we want to investigate how the coupling of quarks to the Polyakov-loop will suppress the distribution of diquark states which arise from strong pairing correlations in quark matter.
To this end we shall consider here the scalar diquark channel as a color antitriplet state and describe it in the framework of a generalized Beth-Uhlenbeck approach. 
Such a treatment was developed for describing excitonic correlations in semiconductor plasmas 
\cite{Zimmermann:1985} and two-nucleon correlations in nuclear matter \cite{Schmidt:1990} before being adapted to the case of mesons in quark matter \cite{Hufner:1994ma} and extended recently to the general case of two-quark correlations in quark matter \cite{Blaschke:2013zaa}.

This approach allows for a microscopic description of the occurrence of bound states in the equation of state of a nonideal plasma and their dissociation at high phase space densities due to the Mott effect.
We develop this approach further by the coupling to the Polyakov loop 
analoguous to Ref.~\cite{Wergieluk:2012gd}, see also \cite{Yamazaki:2012ux,Yamazaki:2013yua}.
Here we will demonstrate how the Polyakov-loop coupling leads to a strong suppression of the colored diquark states in the confining phase in straight analogy to the case of the color triplet quark states.
As a striking elucidation of this effect we consider the superstrong coupling case for which the diquark becomes degenerate in mass to the pion so that the only difference between pion and diquark contributions to the thermodynamics are their numbers of degrees of freedom and the fact that pions are
color neutral while the diquarks form a color antitriplet.

The present work can be considered an important step towards a description of baryons in quark matter.
Any microscopic model which aims at this goal
has to introduce diquarks first as elements in a Faddeev-type description of baryons as three-quark states (see, e.g., \cite{Cahill:1988zi,Cahill:1988bh,Reinhardt:1989rw} for early works and \cite{Wang:2010iu,Blanquier:2011zz} for recent studies at finite temperature and chemical potential).
Such a description has to explain why in the thermodynamics of quark matter with meson-, diquark-, and baryon correlations only mesons and baryons remain as observable degrees of freedom in the confined phase. 
With the present work we provide for the example of the PNJL model the still missing element, the suppression of diquark states in the confining phase.

\section{Mesons and diquarks in PNJL quark matter}


We base the approach on the PNJL model Lagrangian including diquark
interaction channels besides the standard chirally
symmetric scalar-pseudoscalar meson interaction for the isospin symmetric case
($\mu_u=\mu_d=\mu$ and $m_u=m_d=m_0$)
\bea
\mathscr{L}
	&=&
	\bar{q}[\ii\slashed\partial - m_0+\gamma_0(\mu-\ii A_4)]q \nonumber\\
	&&+\mathscr{L}_{\rm int}
	-\mathscr{U}(\Phi, \Bar{\Phi}; T)~,\\
\mathscr{L}_{\rm int}
	&=&
	G_{\rm S}\left[(\bar{q}q)^2+(\bar{q}\ii\gamma_5\tau q)^2\right]\nonumber\\
	&&+ G_{\rm D}\sum_{A=2,5,7}
	(\bar{q} \ii\gamma_5\tau_2\lambda_Aq^c)
	(\bar{q}^c\ii\gamma_5\tau_2\lambda_A q)	~.\nonumber
\eea
Here, $q^c=C\bar{q}^T$ with $C=i \gamma^2\gamma^0$ denote the charge conjugate quark fields, 
$\lambda_A$, $A=2,5,7$, the antisymmetric Gell-Mann matrices in color space and $\tau_i$, $i=1,2,3$
the Pauli matrices in flavor space. $G_S$ and $G_D$ are dimensionful coupling constants.
The Polyakov-loop potential $\mathscr{U}(\Phi, \Bar{\Phi}; T)$ is taken in the
polynomial form  \cite{Ratti:2005jh}, with the parameters taken from that
reference.
The homogeneous gluon background field in the Polyakov gauge is a diagonal
matrix in color space
$A_4=\lambda_3\phi_3+\sqrt{3}\lambda_8\phi_8 =
{\rm diag}(\phi_3+\phi_8,-\phi_3+\phi_8,-2\phi_8)$.
The Polyakov loop field $\Phi$ is defined via the color trace over the gauge-invariant average of the Polyakov line $L(\vec{x})$ \cite{Ratti:2005jh}, which for homogeneous fields becomes
rather simple
\bea
\Phi&=&\frac{1}{N_c}\Tr_c\left[\exp \left(\ii\beta A_4 \right) \right]
\nonumber\\
&=& \frac{1}{N_c}\left[{\rm e}^{\ii\beta(\phi_3+\phi_8)}
+{\rm e}^{-\ii\beta(\phi_3-\phi_8)}+{\rm e}^{-2\ii\beta\phi_8}
 \right]~.
\eea
Its complex conjugate is denoted by $\Bar{\Phi}$.
Starting from the Lagrangian, we perform the usual bosonization by means of
Hubbard-Stratonovich transformations, thus integrating out the quark degrees
of freedom to obtain a path integral representation of the partition function
(and thus the thermodynamical potential $\Omega$) in terms of composite fields,
mesons ($M = \sigma, \vec{\pi}$)
and (anti-)diquarks ($D, \bar{D}=\Delta_A, \Delta_A^*$, $A=2,5,7$),
which in Gaussian approximation can be evaluated in a closed form
\cite{Blaschke:2013zaa}
\bea
\label{partition}
\mathscr{Z}_{\rm Gau{\ss}} &=& 
	\mathscr{Z}_{\rm MF} \Pi_{X=M,D,\bar{D}}\mathscr{Z}_{X}~.
\eea
For the thermodynamic potential $\Omega=-(T/V)\ln \mathscr{Z}$ we obtain
accordingly
\bea
\Omega_{\rm Gau{\ss}}=	\mathscr{U}(\Phi, \Bar{\Phi}; T)
	+\frac{\sigma_{\rm MF}^2}{4G_{\rm S}}
	+	\Omega_{\rm Q}
	+
	\Omega_{\rm M}
	+
	\Omega_{\rm D}
	+
	\Omega_{\rm \bar{D}}
		~,
\eea
with
\bea	
	\Omega_{\rm Q}
	= - \frac{1}{2}\frac{T}{V}\Tr\ln\left[\beta S_{\rm Q}^{-1}\right]
	~,
\label{OmegaQ}
\eea
containing the inverse quark propagator in the mean field approximation
with the Nambu-Gorkov matrix representation 
\bea	
S_{\rm Q}^{-1} &=& \begin{pmatrix}
		(\ii z_n+\hat{\mu})\gamma_0
		-\boldsymbol{\gamma}\cdot{\bf p}
		-m
\hspace{-2mm}		&\Delta_{\rm MF} \ii\gamma_5\tau_2\lambda_2
		\nonumber\\
\hspace{-2mm}		\Delta_{\rm MF}^*\ii\gamma_5\tau_2\lambda_2
&\hspace{-4mm}	(\ii z_n-\hat{\mu})\gamma_0
		-\boldsymbol{\gamma}\cdot{\bf p}
		-m
	\end{pmatrix}.
	\\
\label{propagator}
\eea
Here, $z_n=(2n+1)\pi T$ are the fermionic Matsubara frequencies and
we have introduced the combinations
\bea
m&=&m_0+\sigma_{\rm MF}~, \\
\hat{\mu}&=&\mu -\ii A_4  \nonumber\\
&=& {\rm diag}(\mu-\ii\phi_3-\ii\phi_8,\mu+\ii\phi_3-\ii\phi_8,\mu+2\ii\phi_8)\nonumber\\
&=& {\rm diag}(\mu_r,\mu_g,\mu_b)~.
\eea
In the  present work, we consider both cases, NJL and PNJL, but we will
restrict ourselves to the normal phase without color superconductivity
($\Delta_{MF}=0$).

Note that due to the presence of the diquark fields and the background gauge
field the color trace is not trivial.
In \cite{Blaschke:2013zaa}, we have neglected the gluon background field
(NJL model: $A_4=0$) and performed an expansion w.r.t. the
fluctuations of the composite fields (mesons and diquarks) around their
(homogenous) mean field values up to Gaussian order where the path integral
for the partition function can be evaluated in a closed form.

The functional trace $\Tr$ is defined as a sum over 4-momenta
times the trace $\tr$ over the internal degrees of freedom.
In the infinite volume limit this
becomes
\beq
        \Tr = \sum\limits_{p_n} \tr \;\rightarrow\; V \sum_{z_n}  \int\frac{d^3p}{(2\pi)^3} \tr \,,
\eeq
where $z_n$ denote the Matsubara frequencies, which are fermionic for quarks and baryons,
and bosonic for mesons and diquarks.
The trace in Dirac and flavor spaces in (\ref{OmegaQ}) is readily performed
and after Matsubara summation we arrive for the quark thermodynamical
potential at \cite{Fukushima:2003fw,Ratti:2005jh}
\begin{widetext}
\bea	
\Omega_{\rm Q}
	&=&
	-2N_cN_f\int^\Lambda\frac{d^3p}{(2\pi)^3} E_p
	-2N_f T \int\frac{d^3p}{(2\pi)^3}\left\{
	\tr_{c=r,g,b} \ln\left[1 + {\rm e}^{-(E_p-\mu_c)/T}\right]
	+ \tr_{c=r,g,b} \ln\left[1 + {\rm e}^{-(E_p+\mu_c)/T}\right]\right\}
	~,\nonumber\\
&=&
	-2N_cN_f\int^\Lambda\frac{d^3p}{(2\pi)^3} E_p
	-2N_f T \int\frac{d^3p}{(2\pi)^3}\bigg\{
	\ln\left[\left(1 + Y{\rm e}^{-\ii\beta(\phi_3+\phi_8)}\right)
\left(1 + Y{\rm e}^{\ii\beta(\phi_3-\phi_8)}\right)
\left(1 + Y{\rm e}^{2\ii\beta\phi_8}\right)
\right]\nonumber\\
&&	+ \ln\left[\left(1 + \bar{Y}{\rm e}^{\ii\beta(\phi_3+\phi_8)}\right)
\left(1 + \bar{Y}{\rm e}^{-\ii\beta(\phi_3-\phi_8)}\right)
\left(1 + \bar{Y}{\rm e}^{-2\ii\beta\phi_8}\right)\right]\bigg\}
	~,\nonumber\\
	&=&
	-2N_cN_f\int^\Lambda\frac{d^3p}{(2\pi)^3} E_p
	-2N_f T \int\frac{d^3p}{(2\pi)^3}\left\{
\ln\left[1 + 3\bar{\Phi}Y+3\Phi Y^2 + Y^3\right]
	+ \ln\left[1 + 3{\Phi}\bar{Y}+3\bar{\Phi} \bar{Y}^2 + \bar{Y}^3\right]
\right\}
	~,
\label{Omega_Q}
\eea
\end{widetext}
where we have introduced the abbreviations
$Y={\rm e}^{-(E_p-\mu)/T}$ and $\bar{Y}={\rm e}^{-(E_p+\mu)/T}$.
Removal of the zero-point energy term (``no sea'' approximation) and integration by parts
gives the thermodynamic potential in the form
\bea
\label{Omega_Q-f}
\Omega_{\rm Q}
	&=&
-\frac{2N_c N_f}{3}\int\frac{dp}{2\pi^2}\frac{p^4}{E_p}
	\left[f_\Phi^+(E_p) + f_\Phi^-(E_p)\right] ,
\eea
with the generalized Fermi distribution functions
(cf. Ref.~\cite{Hansen:2006ee})
\bea
\label{f_Phi}
f^+_\Phi(E_p)&=&
\frac{(\bar{\Phi}+2{\Phi}Y)Y+Y^3}{1+3(\bar{\Phi}+{\Phi}Y)Y+Y^3}
~,\nonumber\\
f^-_\Phi(E_p)&=&
\frac{({\Phi}+2\bar{\Phi}\bar{Y})\bar{Y}+\bar{Y}^3}{1+3({\Phi}+\bar{\Phi}\bar{Y})\bar{Y}+\bar{Y}^3}~.
\eea
The limiting cases of the confined phase ($\Phi=\bar{\Phi}=0$) and the deconfined phase 
($\Phi=\bar{\Phi}=1$) these distributions are the Fermi functions
\bea
f^\pm_\Phi(\omega)|_{\Phi=0}&=&\frac{1}{\exp[3(\omega\mp \mu)/T]+1}~,
\\
f^\pm_\Phi(\omega)|_{\Phi=1}&=&\frac{1}{\exp[(\omega\mp \mu)/T]+1}~.
\eea
In the confinement case, Fermi distribution functions with rescaled temperature arise ($T\to T/N_c$), 
so that for a given temperature $T$ exponentially fewer quarks get excited than in the ordinary 
Fermi gas case for $\Phi \to 1$.
The thermodynamic potential of the meson- and diquark channels in Gaussian
approximation is
\bea	
	\Omega_{\rm X}
	=
	 \frac{1}{2}\frac{T}{V}\Tr\ln\left[\beta^2 S_{\rm X}^{-1}\right]~,
	~{\rm X=M,~ D,~\bar{D}}~.
\label{Omega_X}
\eea
where the inverse meson- and diquark propagators take the generic form
\bea	
	S_{\rm X}^{-1}(\ii z_n, {\bf q})
	=\frac{1}{G_{\rm X}} - \Pi_{\rm X}(\ii z_n, {\bf q})
	~,
\label{corr-prop}
\eea
with the polarization functions $\Pi_{\rm X}(\ii z_n, {\bf q})$ defined
in the RPA approximation as one-loop integrals which involve combinations
of quark-quark and quark-antiquark propagators with the
corresponding vertex functions in the meson- and diquark channels,
respectively.
The required polarization loop integrals for mesons and diquarks
are given in Ref.~\cite{Blaschke:2013zaa} for the NJL model.
For the PNJL case, one has to replace in the final expressions the
Fermi functions by the generalized Fermi distribution
functions (\ref{f_Phi}).

For the further evaluation of the thermodynamic potential (\ref{Omega_X}) 
it is convenient to work in a polar representation of the complex propagator function 
which results from analytic continuation of (\ref{corr-prop}) into the complex plane
\bea	
	S_{\rm X}(\omega +i\eta, {\bf q})
	=|S_{\rm X}(\omega, {\bf q})| \exp\left[i\delta_{\rm X}(\omega, {\bf q})\right]
	~,
\label{corr-prop-polar}
\eea
and which defines the phase shift functions $\delta_{\rm X}(\omega, {\bf q})$.
The medium dependence of these functions encodes important physical effects 
such as the merging of the discrete bound state part of the spectrum with the 
continuum of scattering states in the Mott transition. 
The details will be explained later in subsection \ref{ssec:thdyn} when discussing 
the numerical results of the present study. 
An expression for the second virial coefficient in the virial expansion of the 
quantum statistical partition function in terms of medium independent  phase shifts
has been given first by Uhlenbeck and Beth \cite{Beth:1937zz},
who exploited the quantum mechanical models for hard sphere potential scattering.  
In this standard Beth-Uhlenbeck approach which is valid in the low-density limit
one can also employ experimental data on two-particle scattering in free space, 
when they are available as, e.g., in the cases of Coulomb scattering \cite{Ebeling:1968}, 
nucleon-nucleon scattering \cite{Horowitz:2005zv}
or pion nucleon scattering \cite{Weinhold:1997ig}. 

The generalization of the Beth-Uhlenbeck EoS to higher densities has been developed in 
Refs.~\cite{Zimmermann:1985,Schmidt:1990}. 
Out of different aspects of this generalized Beth-Uhlenbeck approach it was in particular 
the modifications of the phase shifts due to a lowering of the continuum threshold 
for scattering states which entails the Mott dissociation of bound states that was interesting
for the application to mesonic correlations in quark matter within the NJL model \cite{Hufner:1994ma}. 
What hindered a broad application of the generalized Beth-Uhlenbeck approach unifying
microscopic approach to the phenomenology of the hadron-to-quark-matter phase transition
was mainly the absence of quark confinement in the NJL model. 
With the advent of the PNJL models (see, e.g.,
\cite{Fukushima:2003fw,Megias:2004hj,Ratti:2005jh,Sasaki:2006ww,Hansen:2006ee,Schaefer:2007pw,Blaschke:2007np,Rossner:2007ik,Radzhabov:2010dd,Horvatic:2010md}) the situation could be amended
since colored quark excitations get suppressed by the coupling to the Polyakov-loop.
In the present work we show how this Polyakov-loop suppression of colored states works for 
the diquark fields which are considered here together with their mesonic counterparts within the 
Polyakov-loop extension of the generalized Beth-Uhlenbeck approach to two-particle states in quark 
matter as derived in Ref.~\cite{Blaschke:2013zaa}.

In the PNJL model the thermodynamic potential for a meson is not directly 
affected by the Polyakov loop and takes the same form as given in \cite{Blaschke:2013zaa},
\bea
\label{GBU_M}
	\Omega_{\rm M}
	&=&
	d_{\rm M}T \int\frac{{\rm d}^3q}{(2\pi)^3}~
	\int_0^\infty\frac{{\rm d}\omega}{2 \pi}~
	\bigg\{
	\ln\left(1-\ee^{-(\omega-\mu_{\rm M})/T}\right)
\nonumber\\
	&&+\ln\left(1-\ee^{-(\omega+\mu_{\rm M})/T}\right)
	\bigg\}
	\frac{d\delta_{\rm M}(\omega, {\bf q})}{d\omega}
	~,
\eea
where $\mu_{\rm M}=\mu_i - \mu_j$ is the chemical potential of a meson $M$ 
composed of quark $i$ with chemical potential $\mu_i$ and antiquark $j$ with 
chemical potential $-\mu_j$. 
Since in the color singlet mesons the colors of quark and antiquark get neutralized,
the meson chemical potential does not contain the gluon background fields $\phi_3$
and $\phi_8$. 
The meson degeneracy factor is denoted by $d_{\rm M}$ and the vacuum 
contribution has been removed.

In such a formulation, the dissociation of the mesonic bound state in a hot, dense
medium by the Mott effect is encoded in the behaviour of the in-medium phase
shift.
The analogous result for the diquark thermodynamic potential is derived 
in the next section for the PNJL model.

\section{Color trace for diquark thermodynamical potential}
\label{app:color}	
To obtain the diquark thermodynamics one starts from the bosonic
thermodynamic potential (\ref{Omega_X}) for the case $X=D$, 
taking the form (\ref{GBU_M}) where the meson chemical potential $\mu_M$
has to be replaced by the diquark chemical potentials $\mu_A$ for the three
diquark channels $D_A,~A=2,5,7$, which are diagonal matrices in color space
\bea
\mu_{2}&=&\mu_r + \mu_g = 2\mu -2\ii\phi_8\nonumber\\
\mu_{5}&=&\mu_r + \mu_b = 2\mu -\ii(\phi_3-\phi_8)\nonumber\\
\mu_{7}&=&\mu_r + \mu_g = 2\mu +\ii(\phi_3+\phi_8)~.
\eea
In the diquark thermodynamic potential which otherwise has the form similar
to the pion one, the remaining color trace is evaluated in the following way
\begin{widetext}
\bea	
\Omega_{\rm D}
	&=&
	\int\frac{d^3q}{(2\pi)^3}\int\frac{d\omega}{2\pi}
	\left\{3\omega
	+T\tr_{A=2,5,7} \ln\left[1 - {\rm e}^{-(\omega-\mu_A)/T}\right]
+T\tr_{A=2,5,7} \ln\left[1 - {\rm e}^{-(\omega+\mu_A)/T}\right]\right\}
\frac{d\delta_D(\omega)}{d\omega}	~,\nonumber\\
&=&
	\int\frac{d^3q}{(2\pi)^3}\int\frac{d\omega}{2\pi}\bigg\{3\omega
	+T\ln\left[\left(1 - X{\rm e}^{-2\ii\beta\phi_8}\right)
\left(1 - X{\rm e}^{-\ii\beta(\phi_3-\phi_8)}\right)
\left(1 - X{\rm e}^{\ii\beta(\phi_3+\phi_8)}\right)
\right]\nonumber\\
&&	+T \ln\left[\left(1 - \bar{X}{\rm e}^{2\ii\beta\phi_8}\right)
\left(1 - \bar{X}{\rm e}^{\ii\beta(\phi_3-\phi_8)}\right)
\left(1 - \bar{X}{\rm e}^{-\ii\beta(\phi_3+\phi_8)}\right)\right]\bigg\}
\frac{d\delta_D(\omega)}{d\omega}	~,\nonumber\\
	&=&
	\int\frac{d^3q}{(2\pi)^3}\int\frac{d\omega}{2\pi}\left\{3\omega
+T\ln\left[1 - 3{\Phi}X+3\bar{\Phi} X^2 - X^3\right]
	+T \ln\left[1 - 3\bar{\Phi}\bar{X}+3{\Phi} \bar{X}^2 - \bar{X}^3\right]
\right\}\frac{d\delta_D(\omega)}{d\omega}
	~,
\label{Omega_D}
\eea
\end{widetext}
where we have introduced the abbreviations
$X={\rm e}^{-(\omega-2\mu)/T}$, $\bar{X}={\rm e}^{-(\omega+2\mu)/T}$
and dropped the explicit notation of the three-momentum {\bf q} as an argument 
of the phase shifts. This shorthand notation we shall use from now on.
Removal of the zero-point energy term (``no sea'' approximation) 
results in the diquark thermodynamic potential 
\bea
\label{GBU_D}
	\Omega_{\rm D}
	&=&
	T\int\frac{{\rm d}^3q}{(2\pi)^3}~
	\int_0^\infty\frac{{\rm d}\omega}{2 \pi}~
	\bigg\{ \ln\left[1-3({\Phi}-\bar\Phi X)X-X^3\right]
	\nonumber\\
	&&+\ln\left[1-3(\bar{\Phi}-{\Phi}\bar{X})\bar{X} - \bar{X}^3\right]
	\bigg\}
	\frac{d\delta_{\rm D}(\omega)}{d\omega} .
\eea
While the mesons are color neutral the scalar diquarks are color antitriplet states 
and thus ``see'' the Polyakov loop.
Note that in the confining phase it holds that $\Phi=\bar{\Phi}=0$ which removes 
the contributions from one- and two- diquark states so that in this phase only the coherent
sum of three diquarks with complementary colors can propagate.
This effect is in full analogy to the case of colored quark excitations which are also
suppressed by the Polyakov-loop factors in the confinement phase,
see Eq. (\ref{Omega_Q}) \cite{Fukushima:2003fw,Ratti:2005jh}.

After integration by parts (\ref{GBU_D}) takes the form
\bea
\label{Omega_D-g}
\Omega_{\rm D} &=&
-3 \int\frac{d^3p}{(2\pi)^3}\hspace{-2mm}\int\frac{d\omega}{2\pi}
	\left[g_\Phi^+(\omega) + g_\Phi^-(\omega)\right] \delta_D(\omega),
	\nonumber\\
\eea
with the generalized Bose distribution functions
\bea
\label{g_Phi}
g^+_\Phi(\omega)&=&
\frac{({\Phi}-2\bar{\Phi}{X}){X}+{X}^3}{1-3({\Phi}-\bar{\Phi}{X}){X}-{X}^3}~,\nonumber\\
g^-_\Phi(\omega)&=&
\frac{(\bar{\Phi}-2{\Phi}\bar{X})\bar{X}+\bar{X}^3}{1-3(\bar{\Phi}-{\Phi}\bar{X})\bar{X}-\bar{X}^3}~.
\eea

In the limiting cases of the confined phase ($\Phi=\bar{\Phi}=0$) and the 
deconfined phase ($\Phi=\bar{\Phi}=1$) these functions 
go over to the Bose functions
\bea
g^\pm_\Phi(\omega)|_{\Phi=0}&=&\frac{1}{\exp[3(\omega\mp 2\mu)/T]-1}~,
\\
g^\pm_\Phi(\omega)|_{\Phi=1}&=&\frac{1}{\exp[(\omega\mp 2\mu)/T]-1}~,
\eea
which show that in the confined phase the thermal excitation of diquarks and antidiquarks 
is suppressed by a similar mechanism as the excitation of quarks and antiquarks.

This is the main new result of this work. In the following section we will
obtain numerical results for it and discuss its consequences.
\begin{figure}[!htb]
\includegraphics[width=0.47\textwidth]{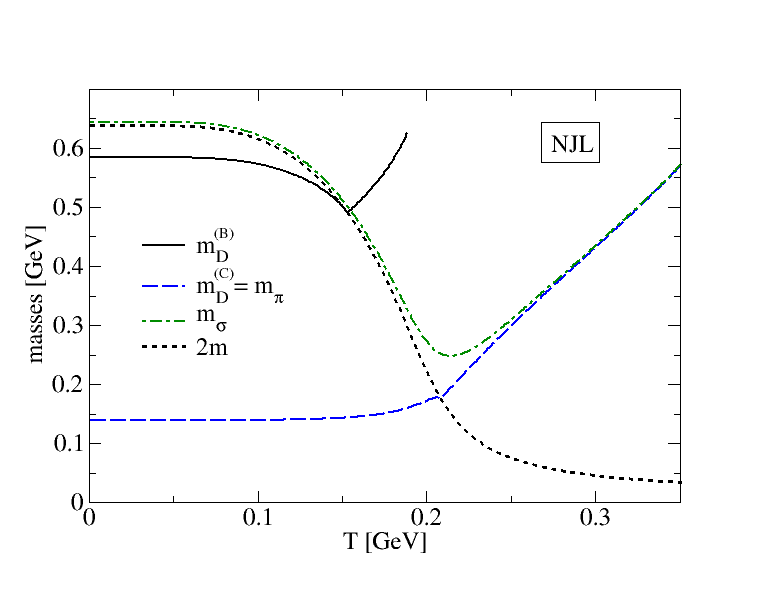}
\includegraphics[width=0.47\textwidth]{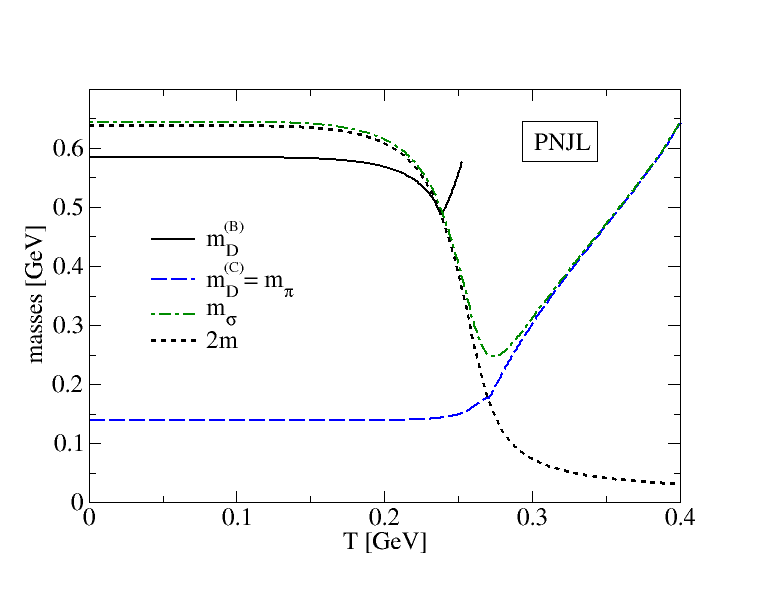}
\caption{Mass spectrum of pions, sigma mesons and diquarks as
functions of the temperature $T$.
The diquark mass $m_D^{(B)}$ ($m_D^{(C)}$) corresponds to the case of strong 
(superstrong) diquark coupling.
Also shown is the relevant threshold $2m$ for mesons and diquarks.
Upper panel: without Polyakov-loop; Lower Panel: with Polyakov-loop.}
\label{fig:masses1}
\end{figure}

\section{Results}

The parameters employed for the numerical studies are a bare quark mass
$m_0 = 5.5$~MeV, a three-momentum cutoff $\Lambda = 639$~MeV and a scalar coupling constant
$G_{\rm S}\Lambda^2 = 2.134$.
We consider three values for the diquark coupling constant:
(A) Fierz value: $G_D/G_S=3/4$ , (B) strong coupling $G_D/G_S=1.0$ and
(C) superstrong coupling: $G_D/G_S=1.5$.
With these parameters one finds in vacuum a constituent quark mass of
319~MeV, a pion mass of 138~MeV and pion decay constant $f_\pi=92.4$ MeV.
The vacuum mass of the  $\sigma$-meson is 644~MeV,
which is thus slightly unbound.
The scalar diquark is unbound for the Fierz value of the coupling (case A)
and bound for the strong and superstrong diquark couplings ($m_D^{(B)}=582$ MeV for cases B 
and $m_D^{({C})}=m_\pi$ for case C, resp.).
In the present work, we restrict ourselves to applications at finite temperatures
and vanishing chemical potential in the isospin-symmetric case of two-flavor
quark matter.

\subsection{Mass spectrum in the NJL/PNJL model at finite temperature}

As a first step, the mean field gap equation for the quark mass is solved
as a function of temperature for the NJL and PNJL models,
respectively. These results serve as inputs for solving the Bethe-Salpeter
equations for mesons and diquarks in the medium.
The masses of quarks, pions, $\sigma$ mesons and diquarks are obtained as
poles of their propagators and are shown in Fig.~\ref{fig:masses1}.
We observe that the chiral symmetry restoration which is reflected in the
dropping quark mass function, induces a Mott effect for the pion and the
scalar diquark. For both states the kernel of their Bethe-Salpeter equation
contains a Pauli blocking term since they are composed of two fermions.
This Pauli blocking partly compensates the effect of the quark selfenergies
(dropping masses) and leads to a stabilization of the bound state
masses against medium effects. This results in the crossing of the
bound state masses with the continuum threshold, leading to the dissociation
of these bound states.
\subsection{Phase shifts of mesons and diquarks in quark matter}
\label{ssec:thdyn}
\begin{figure}[!htb]
\includegraphics[width=0.42\textwidth]{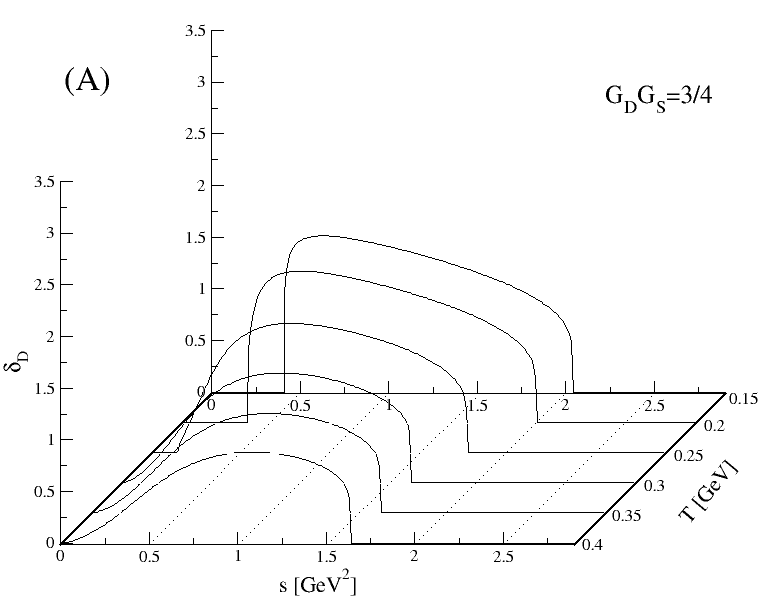}
\includegraphics[width=0.42\textwidth]{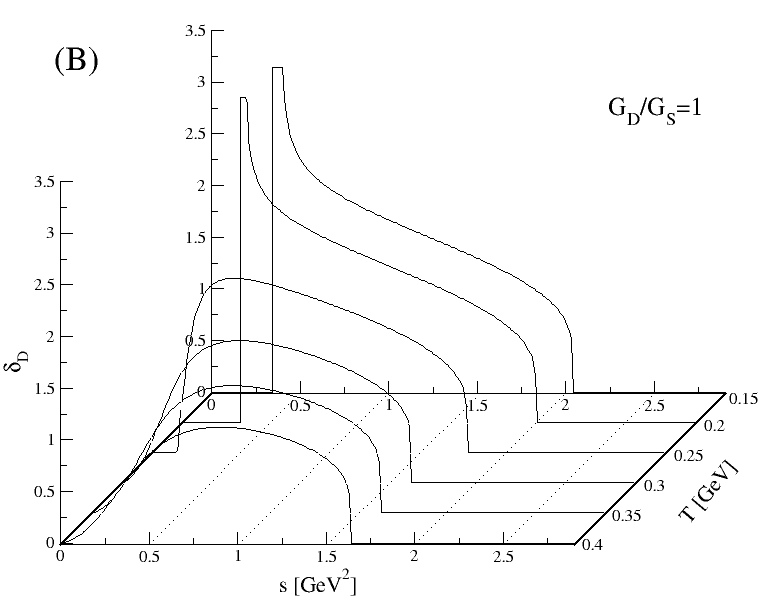}
\includegraphics[width=0.42\textwidth]{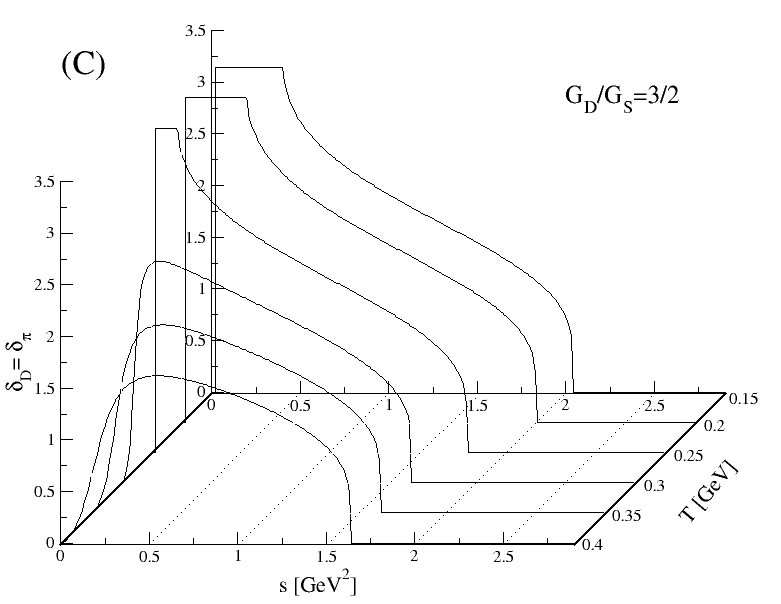}
\caption{
Diquark phase shift $\delta_D$ as a function of squared center of mass energy
$s$ at different temperatures from $T=150$ MeV to $400$ MeV for the three cases
of diquark coupling:
(A) Fierz value: $G_D/G_S=3/4$ , (B) strong coupling $G_D/G_S=1.0$ and
(C) superstrong coupling: $G_D/G_S=1.5$.
The jump of the phase shift from $0$ to $\pi$ indicates the position of a bound state
in the spectrum below the threshold of the continuum states which is situated where the
phase shift starts decreasing towards zero.
With increasing temperature the threshold moves to lower $s$-values and the phase shift
jumps from $\pi$ to $0$ when the Mott temperature is reached where the bound
state gets dissociated. This behaviour is in accordance with the Levinson theorem.
}
\label{fig:PHD}
\end{figure}

In the following, we will discuss the results for the diquark phase shifts
(see also Refs. \cite{Blaschke:2013zaa}) and their consequences for the
thermodynamics of quark-meson-diquark matter at finite temperature, with and
without the coupling to the Polyakov-loop.
The solution for the diquark phase shifts at finite temperature ($\mu=0$) is
shown in Fig.~\ref{fig:PHD} as a function of the invariant mass variable $s$
for different temperatures.
Here we made the simplifying assumption that, even in the medium, the phase shifts 
are Lorentz invariant and identify the function $\delta_X(s)$ with $\delta_X(\omega=\sqrt{s})$ 
calculated at rest ({\bf q}=0) for given temperature and chemical potential of the medium.
The bound state mass is located at the jump of the phase shift from $0$ to
$\pi$ and this jump corresponds to a delta-function in the Beth-Uhlenbeck
formulas  (\ref{GBU_M}) and (\ref{GBU_D})
for the correlations. In the case when the continuum of the
scattering states can be neglected since it is separated by a sufficient
energy gap from the bound state, we obtain the limiting case of thermodynamics
of a statistical ensemble of on-shell correlations.

\subsection{Beth-Uhlenbeck equation for mesons and diquarks in quark matter}

Now we want to study the thermodynamics of the meson and diquark correlations
in a hot and dense medium encoded in the thermodynamic potentials (\ref{GBU_M})
and (\ref{GBU_D}), respectively.

\begin{figure}[!htb]
\includegraphics[width=0.43\textwidth]{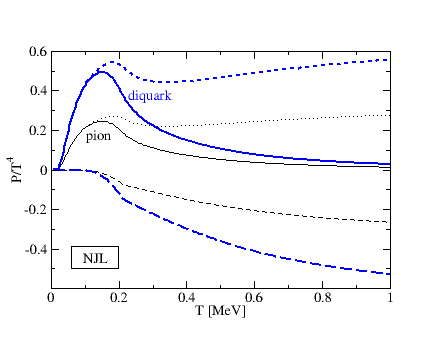}
\includegraphics[width=0.43\textwidth]{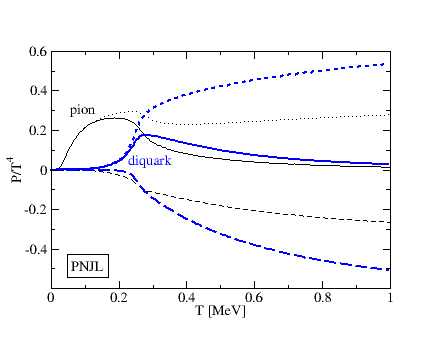}
\includegraphics[width=0.43\textwidth]{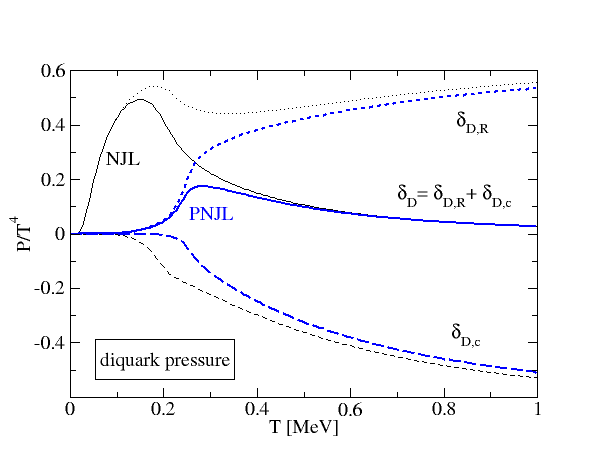}
\caption{Contributions from the resonant (positive) and scattering continuum (negative) parts to the
total pressure (thick lines) for diquarks and pions in the NJL model (upper panel)
and in the PNJL model (middle panel) as a function of temperature.
In the lower panel a direct comparison is made of the diquark pressure with its resonant and continuum
contributions for the NJL and the PNJL model, respectively, which demonstrates the strong suppression of the diquark pressure due to the Polyakov-loop coupling in the confining phase at low temperatures.
In this figure we have chosen the superstrong coupling case (C) for which
the masses of pions and diquarks are degenerate.
}
\label{fig:p_D}
\end{figure}

In Fig.~\ref{fig:p_D} we show the pressure as a function of the temperature for
pions and diquarks within the NJL and the PNJL models, respectively.
Let us note that within NJL-type models a general decomposition of the
phase shifts into a resonant (R) and a continuum (c) part can be made
\bea
\label{split}
\delta_{\rm X}(\omega)=
\delta_{\rm X,R}(\omega)+\delta_{\rm X,c}(\omega)~,
\eea
where both parts are uniquely defined by the propagator of the correlation
(see Ref.~\cite{Blaschke:2013zaa}) and are important to establish accordance with the
Levinson theorem in medium \cite{Wergieluk:2012gd,Dubinin:2013yga}
\bea
\label{Levinson}
\int_0^\infty d\omega \frac{d \delta_{\rm X}(\omega)}{d\omega}&=&0~.
\eea
The more conventional form of the Levinson theorem \cite{Dashen:1969ep}
introduces the energy level of the continuum threshold $\omega_{\rm thr}=\sqrt{q^2+m_{\rm thr}^2}$,
where $m_{\rm thr}=2m$ applies for the continuum of two-particle states (mesons, diquarks) composed 
of quarks with equal mass $m$
\bea
\label{Levinson1}
\int_0^{\omega_{\rm thr}} d\omega \frac{d \delta_{\rm X}(\omega)}{d\omega}
&=& - \int_{\omega_{\rm thr}}^\infty d\omega \frac{d \delta_{\rm X}(\omega)}{d\omega}
\nonumber\\
&=& \delta_{\rm X}(\omega_{\rm thr}) - \delta_{\rm X}(\infty)~.
\eea
Since below the threshold can be only a discrete number $n_{\rm B,X}$ of bound states in the channel
${\rm X}$, each contributing an amount of $\pi$ to the change in the phase shift 
at the bound state energies $\omega_{\rm X, i}=\sqrt{q^2+m_{\rm i}^2}$ with ${\rm i}=1,\dots, n_{\rm B,X}$, 
it follows the Levinson theorem in the form
\bea
\label{Levinson2}
\pi ~n_{\rm B,X} = \delta_{\rm X}(\omega_{\rm thr}) - \delta_{\rm X}(\infty)~,
\eea
which applies also in the case of a hot and dense medium.

In particular, when due to the chiral restoration the dropping mass of the quarks entails a lowering of the continuum threshold $\omega_{\rm thr}$ which triggers 
the dissolution of the bound states into the continuum (the Mott effect), then 
$n_{\rm B,X}=0$. 
In that case, it holds that  $\delta_{\rm X}(\omega_{\rm thr}) = \delta_{\rm X}(\infty)$,
which would be strongly violated if the phase shift  (\ref{split}) would be approximated by the resonance
part only. This can be demonstrated, e.g., by employing a Breit-Wigner type ansatz for the phase shift
\bea
\label{Breit-Wigner}
\delta_{\rm X,R}(\omega) = \frac{\pi}{2} 
+ {\rm arctan} \left(\frac{\omega-\omega_{\rm X,R}}{\Gamma_{\rm X}}\right)~,
\eea 
which would yield
\bea
\label{Breit-Wigner}
\delta_{\rm X,R}(\omega_{\rm thr}) -  \delta_{\rm X,R}(\infty) &=& 
\arctan \left(\frac{\omega_{\rm thr}-\omega_{X,R}}{\Gamma_X} \right) - \frac{\pi}{2} 
\nonumber\\
&\neq& 0~.
\eea 
This violation of the Levinson theorem elucidates the importance of continuum background contribution to the phase shift (\ref{split}).

In order to make the role of the Polyakov-loop suppression most explicit we have chosen 
{in Fig.~\ref{fig:p_D}} 
the superstrong coupling case with the parametrization $G_D=1.5~G_S$, 
for which the masses of pions and diquarks are degenerate. 
This is similar to the mass spectrum of two-color QCD \cite{Ratti:2004ra,Strodthoff:2011tz,Zablocki:2012zz,Strodthoff:2013cua}.
{
It is this case for which the main result of this paper can be most clearly demonstrated, namely} 
that the coupling to the Polyakov loop very effectively suppresses the partial pressure of the 
colored diquark states,  thus making the thermodynamics of the PNJL model more ``realistic''
than that of the NJL model.

\begin{figure}[!htb]
\includegraphics[width=0.49\textwidth]{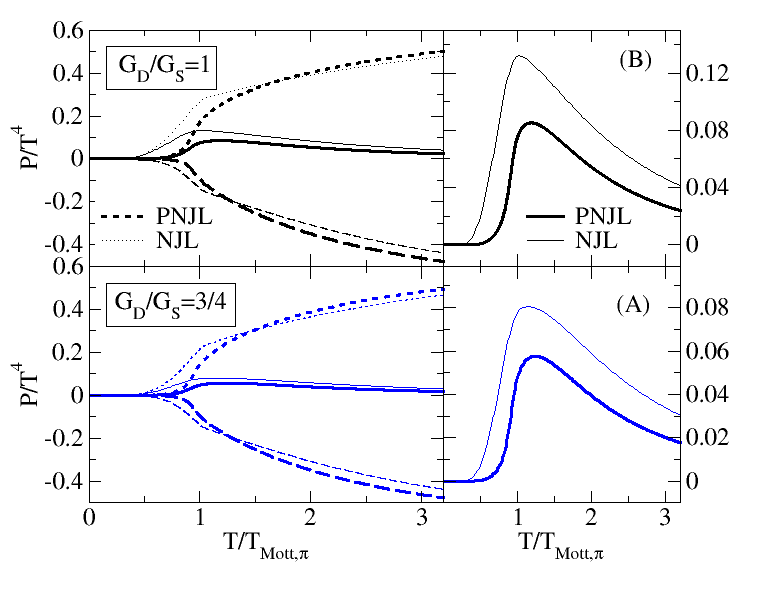}
\caption{Contributions to the diquark pressure in the NJL model (thin lines)
compared to that of the PNJL model (thick lines) as a function of temperature:
resonance (dotted lines), continuum (dashed lines) and total (solid lines)
contributions for two cases of diquark coupling: strong (case (B), upper panels)
and Fierz value (case (A), lower panels).
The right panels demonstrate the suppression of the diquark pressure due to
the Polyakov-loop coupling by comparing the NJL with the PNJL model results.
The diquark pressure contains contributions from both, diquarks and
antidiquarks.}
\label{fig:p_Dzoom}
\end{figure}

When we reduce the diquark coupling to the more realistic case (B), where
the diquark mass is just below the continuum threshold $2m$ and case (A) where
it is even above the threshold, the diquark pressure is lowered further due
to the increase in the diquark mass relative to case (C).
In Fig.~\ref{fig:p_Dzoom} we show the corresponding contributions to the
diquark pressure as a function of the temperature for the NJL and the PNJL model,
respectively.

\begin{figure}[!htb]
\vspace{5mm}
\includegraphics[width=0.49\textwidth]{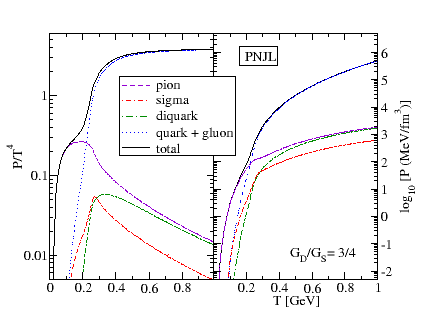}
\caption{Pressure of  {quark-gluon} matter with meson and diquark correlations in the
PNJL model as function of temperature (solid lines).
Also shown are the partial pressures of quarks {and gluons} (dotted lines),
diquarks (long-dash-dotted lines), sigma mesons (dash-dotted lines) and pions
(dashed lines).
The left panel shows the pressure contributions divided by $T^4$
(dimensionless) while on the right panel they are in units of MeV/fm$^3$.
}
\label{fig:p_pidi}
\end{figure}

We are now in a state to summarize the results for the thermodynamics of the
quark-meson-diquark plasma within the PNJL model.
In Fig.~\ref{fig:p_pidi} we show the contributions from quarks, sigma mesons,
pions and diquarks to the total pressure in hot quark matter for case (A), the
Fierz value of the diquark coupling.
We observe that for temperatures up to 
{$T\sim 120$ MeV} 
the total pressure is that of a pion gas while for temperatures exceeding $T\sim 250$ MeV the thermodynamics is that of a quark-gluon plasma. 
The intermediate temperature region is that of the chiral restoration and deconfinement transition, 
where the system consists of a mixture of partons and pions, while heavier hadrons like the sigma 
meson as well as the diquarks are subdominant.
In this transition region the composition of the thermodynamic system is governed by 
three suppression mechanisms for states: 
due to their mass, their spectral broadening (dissociation) and the Polyakov loop. While the mass suppresses states only at low temperatures, the spectral broadening acts for all composite states and suppresses them at high temperatures. 
The colored states (quarks and diquarks) are suppressed in the confined phase where the traced 
Polyakov loop is close to zero. 

This result appears to capture the characteristic features of QCD thermodynamics
at finite temperatures as it is simulated on the lattice \cite{Borsanyi:2010cj,Bazavov:2014pvz}.
However, before a quantitative comparison with lattice QCD can be attempted, more hadronic
states need to be implemented as, e.g., in the recent phenomenological model
\cite{Turko:2011gw,Turko:2013taa,Turko:2014jta,Blaschke:2015nma}.
Another important aspect is the consistent inclusion of correlation effects into the quasiparticle picture
which is under way \cite{Blaschke:2015bxa}.

\section{Conclusions}			
A main result of this work is the derivation of the generalized distribution function for
color SU(3) diquarks in a Polyakov-loop background field.
In the limit of deconfinement, it goes over to the ordinary Bose distribution function while in the opposite case it is responsible for a strong suppression of the colored diquark state.
This has been strikingly demonstrated by considering the superstrong coupling case $G_D=3/2~G_S$ for
which the pion and diquark Bethe-Salpeter equations become degenerate and produce therefore the same
solutions, masses and also phase shifts.
In this case it is just the generalized Bose distribution function which leads to a strong suppression of the partial pressure of the colored diquark states relative to that of the color neutral pions.
We have evaluated the composition (partial pressures) of quark-meson-diquark matter in the two-flavor
PNJL model as a function of the temperature at zero baryon density with the result that in the confinement phase the system becomes a pion gas where quarks and diquarks as colored degrees of freedom are suppressed by the Polyakov-loop and sigma mesons by their mass.
With increasing temperature the system undergoes the transition to the deconfined phase which coincides with chiral symmetry restoration. 
The meson and diquark contributions to the pressure are vanishing due to their dissolution in the Mott transition while their constituents become the dominant component, now forming a Fermi gas of colored states not suppressed by the Polyakov-loop since we are in the deconfined phase.

\section{Acknowledgement}			

This work was supported by the Deutsche Forschungsgemeinschaft (DFG) under
contract BU 2406/1-1 and by the Polish National Science Centre within the ``Maestro'' programme under
contract UMO-2011/02/A/ST2/00306.
The work of D.B. was supported in part by the Hessian LOEWE initiative through HIC for FAIR.
A.D. acknowledges a grant from the Institute for Theoretical Physics of the University of Wroclaw
under contract No. 2470/M/IFT/14 and support by the Bogoliubov-Infeld programme for scientific collaboration between Polish Institutions and the JINR Dubna.



\end{document}